\documentclass[preprint,showpacs,preprintnumbers,amsmath,amssymb]{revtex4}

\usepackage{graphicx}
\usepackage{dcolumn}
\usepackage{bm}

\begin{document}

\preprint{APS/123-QED}

\title{ Influence of charge asymmetry and isospin dependent cross-section on nuclear stopping\\}

\author{Anupriya Jain}
\author{Suneel Kumar}%
 \email{suneel.kumar@thapar.edu}
\affiliation{%
School of Physics and Material Science, Thapar University, Patiala-147004, Punjab (India)\\
}%

\author{Rajeev K. Puri}
\affiliation{
Department of Physics, Panjab University, Chandigarh-160014, India\\
}%
\date{\today}

\begin{abstract}
Using the isospin dependent quantum molecular dynamics model, we study the effect of charge asymmetry and isospin dependent cross-section on nuclear stopping and multiplicity of free nucleons and LMF's. Simulations were carried out for the reactions $^{124}X_{m}+^{124}X_{m}$, where m varies from 47 to 59 and for $^{40}Y_{n}+^{40}Y_{n}$, where n varies from 14 to 23. Our study shows that nuclear stopping as well as the production of LMF's depend strongly on the isospin of the cross-section. 
\end{abstract}

\maketitle
\baselineskip=1.5\baselineskip\

Heavy ion collisions (HIC) provide an unique opportunity to produce small amount of nuclear matter with high density and high temperature in a controlled fashion. By measuring the final products of the collisions, it is possible to learn about the fundamental properties of the hot and compressed nuclear matter namely, the nuclear equation of state (EOS) \cite{1,2,3}. In order to link the experimental observations and equation of state extracted from the heavy ion collisions, we need transport model where nucleon-nucleon two body collisions and mean field effects are carefully treated \cite{4,5,6}.\\ 
One of the key observables in heavy ion collisions (HIC) is the nuclear stopping that can be studied with the help of rapidity distribution \cite{7} and asymmetry of the nucleon momentum distribution \cite{8}. W. Bauer \cite{9} pointed out that nuclear stopping power in intermediate energy HIC, is determined by both mean field as well as in-medium nucleon-nucleon cross-section. It is worth mentioning that symmetry potential was not included in their analysis. 
G. Peilert {\it et al.,} \cite{10} suggested that the degree of approaching isospin equilibration provides a mean to probe the mechanism and power of the nuclear stopping in heavy ion collisions.\\ 
Fen Fu {\it et al.,} \cite{11} calculated both the radial flow and  degree of nuclear stopping using the reactions of Pb + Pb and Ni + Ni at 0.4, 0.8 and 1.2 GeV/nucleon. They found that the expansion velocity as well as the degree of nuclear stopping are higher in heavier system irrespective of the incident energies.\\
Qingfeng Li and Zhuxia Li \cite{12} studied the dependence of nuclear stopping $\langle$ $Q_{ZZ}/A$ $\rangle$ and $\langle$ R $\rangle$  in intermediate energy heavy-ion collisions on system size, initial N/Z, isospin symmetry potential and medium corrections of two-body cross-sections. They showed that the effect of initial N/Z ratio as well as of isospin symmetry potential is weak on stopping. The excitation function of
$\langle$ $Q_{ZZ}/A$ $\rangle$  and $\langle$ R $\rangle$  however, depends on the form of the medium corrections of two-body cross-sections and on the equation of state of nuclear matter (EOS). Moreover, they showed that the behavior of excitation function of $\langle$ $Q_{ZZ}/A$ $\rangle$ and R can provide clearer information about the isospin dependence of the medium correction of two-body cross-sections.\\
Jian-Ye Liu {\it et al.,} \cite{13} studied the nuclear stopping for various colliding systems with different neutron-proton ratios over large domains of incident energy. Nuclear stopping was found to be very sensitive towards the isospin contant of in-medium nucleon-nucleon cross- section above Fermi energy. The results were, however, insensitive towards the symmetry potential. They proposed that nuclear stopping can be used as a new probe to extract the information about the isospin dependence of the in-medium nucleon-nucleon cross-section in intermediate energy heavy ion collisions.\\
In this study, our aim is to pin down the influence of charge asymmetry as well as of different cross-sections ($\sigma_{iso}$ (isospin dependent) and  $\sigma_{noiso}$ (isospin independent)) on nuclear stopping observables like $\langle$ R $\rangle$, $\langle$ $1/Q_{ZZ}$ $\rangle$   and multiplicity of free nucleons and LMF's.\\ 
Our study is performed within the framework of IQMD \cite{5,14} model where hadrons propagate with Hamilton equations of motion:
\begin{equation}
\frac{d{r_i}}{dt}~=~\frac{d\it{\langle~H~\rangle}}{d{p_i}}~~;~~\frac{d{p_i}}{dt}~=~-\frac{d\it{\langle~H~\rangle}}{d{r_i}},
\end{equation}
with
\begin{eqnarray}
\langle~H~\rangle&=&\langle~T~\rangle+\langle~V~\rangle\nonumber\\
&=&\sum_{i}\frac{p_i^2}{2m_i}+
\sum_i \sum_{j > i}\int f_{i}(\vec{r},\vec{p},t)V^{\it ij}({\vec{r}^\prime,\vec{r}})\nonumber\\
& &\times f_j(\vec{r}^\prime,\vec{p}^\prime,t)d\vec{r}d\vec{r}^\prime d\vec{p}d\vec{p}^\prime .
\end{eqnarray}
 The baryon-baryon potential $V^{ij}$, in the above relation, reads as:
\begin{eqnarray}
V^{ij}(\vec{r}^\prime -\vec{r})&=&V^{ij}_{Skyrme}+V^{ij}_{Yukawa}+V^{ij}_{Coul}+V^{ij}_{sym}\nonumber\\
&=& \left [t_{1}\delta(\vec{r}^\prime -\vec{r})+t_{2}\delta(\vec{r}^\prime -\vec{r})\rho^{\gamma-1}
\left(\frac{\vec{r}^\prime +\vec{r}}{2}\right) \right]\nonumber\\
& & +~t_{3}\frac{exp(|\vec{r}^\prime-\vec{r}|/\mu)}{(|\vec{r}^\prime-\vec{r}|/\mu)}~+~\frac{Z_{i}Z_{j}e^{2}}{|\vec{r}^\prime -\vec{r}|}\nonumber\\
& & + t_{6}\frac{1}{\varrho_0}T_3^{i}T_3^{j}\delta(\vec{r_i}^\prime -\vec{r_j}).
\label{s1}
\end{eqnarray}
Here $Z_i$ and $Z_j$ denote the charges of $i^{th}$ and $j^{th}$ baryon, and $T_3^i$, $T_3^j$ are their respective $T_3$
components (i.e. 1/2 for protons and -1/2 for neutrons). Meson potential consists of Coulomb interaction only.
The parameters $\mu$ and $t_1,.....,t_6$ are adjusted to the real part of the nucleonic optical potential.\\ 
The binary nucleon-nucleon collisions are included by employing the collision term of well known VUU-BUU equation. During the propagation, two nucleons are
supposed to suffer a binary collision if the distance between their centroids
\begin{equation}
|r_i-r_j| \le \sqrt{\frac{\sigma_{tot}}{\pi}}, \sigma_{tot} = \sigma(\sqrt{s}, type),
\end{equation}
"type" denotes the ingoing collision partners (N-N, N-$\Delta$, N-$\pi$,..). In addition,
Pauli blocking (of the final
state) of baryons is taken into account by checking the phase space densities in the final states.\\
Nuclear stopping is investigated using three observables. The first one is anisotropy ratio $\langle$ R $\rangle$ \cite{13,14}, defined as: 
\begin{equation}
\langle R \rangle= \frac{2}{\pi}\frac{\left(\sum_{i}p_{\perp}(i)\right)}{\left(\sum_{i}p_{\parallel}(i)\right)}
\end{equation}
where $p_{\perp}(i)= \sqrt{(p_x^2(i) + p_y^2(i)}$ and $p_{\parallel}(i)= p_{Z}(i)$ respectively. If $\langle$ R $\rangle$ =1, then it means complete stopping.\\
Second parameter is the quadrupole moment $\langle$ $Q_{ZZ}$ $\rangle$   \cite{13,14}, defined as:

\begin{equation}
\langle Q_{ZZ} \rangle=\sum_{i}2p_z^2(i)-p_x^2(i)-p_y^2(i)
\end{equation}

For complete stopping $\langle$ $Q_{ZZ}$ $\rangle$ =0.

The third parameter is rapidity distribution Y(i) \cite{14,16}, defined as:

\begin{equation}
Y(i)= \frac{1}{2}ln\frac{E(i)+p_{z}(i)}{E(i)-p_{z}(i)},
\end{equation}

Where, E(i) and $p_{Z}(i)$ are the total energy and longitudinal momentum, respectively.\\
 For the present analysis, simulations are carried out for two set of reactions using soft equation of state. For the first case, mass of the colliding nuclei is fixed to be 40 units, but charge varies from 14 to 23. In other words, we study $^{40}X_{m}+^{40}X_{m}$, where $^{40}X_{m}$ = ($^{40}V_{23}$, $^{40}Sc_{21}$, $^{40}Ca_{20}$, $^{40}Ar_{18}$, $^{40}Cl_{17}$, $^{40}S_{16}$, $^{40}P_{15}$ and $^{40}Si_{14}$ ), respectively. For the second set, we have chosen those reactions where mass of the each colliding nuclei is fixed to be 124 units, but charge varies from 47 to 59. The Second set of the reactions taken are $^{124}Y_{n}+^{124}Y_{n}$, where $^{124}Y_{n}$ = ($^{124}Ag_{47}$, $^{124}Cd_{48}$,  $^{124}In_{49}$, $^{124}Sn_{50}$, $^{124}I_{53}$, $^{124}Cs_{55}$, $^{124}Ba_{56}$ and $^{124}Pr_{59}$ ), respectively.
\begin{figure}
\hspace{-2.0cm}\includegraphics[scale=0.45]{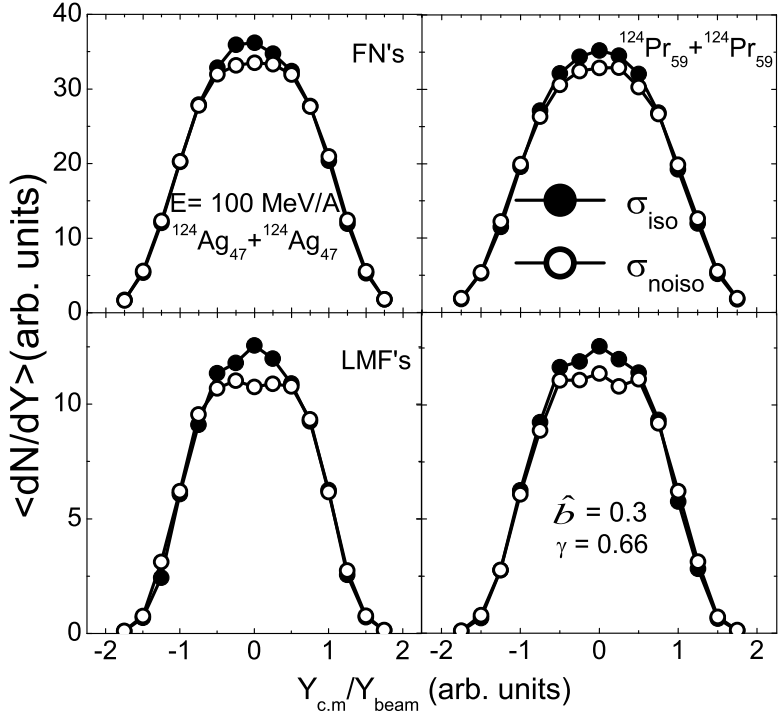}
\caption{\label{Fig:1} Rapidity distribution of FN's and LMF's.}
\end{figure}
Fig.1 shows the rapidity distribution $\langle$ dN/dY $\rangle$ for the emission of free nucleons and LMF's at incident energy of E = 100 MeV/nucleon. To check the role of different cross-sections on the rapidity distribution, two reactions $^{124}Ag_{47}+^{124}Ag_{47}$ and $^{124}Pr_{59}+^{124}Pr_{59}$ are displayed. This choice of reaction panels will throw light on the role of charge asymmetry. As noted by many authors, free nucleons and LMF's are produced from the participant zone whereas IMF's are mostly produced out of the spectator matter.\\
We also noted clear isospin effects on the production of free nucleons and LMF's in the energy region of mid-rapidity. This happens due to the fact that $\sigma_{iso}$ (near mid rapidity) will enhance the binary collisions that results in enhanced production. This effect should diminish as we move away from the mid-rapidity region where either target like or projectile like process dominates.\\
Strikingly, a very little influence (less than $3\%$) is noted due to charge asymmetry is seen. Note that while mass remains fixed, charge of the colliding nuclei varies from 47 units to 59 units.
\begin{figure}
\hspace{-2.0cm}\includegraphics[scale=0.45]{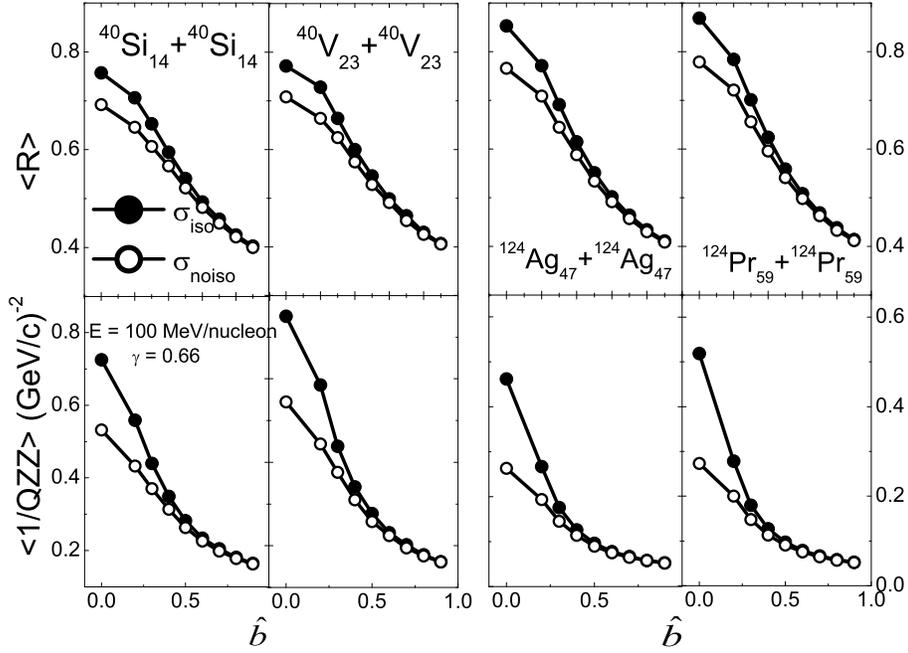}
\caption{\label{Fig:2} Variation of $\langle$ R $\rangle$ and $\langle$ $1/Q_{ZZ}$ $\rangle$ as a function of $\hat{b}$.}
\end{figure}
To check the effect of isospin dependence of cross-section on nuclear stopping, we display in Fig.2 the impact parameter dependence of the stopping observables ($\langle$ R $\rangle$ and $\langle$ $1/Q_{ZZ}$ $\rangle$). The results are displayed at 100 MeV/nucleon for the reactions of $^{40}Si_{14}+^{40}Si_{14}$ and $^{40}V_{23}+^{40}V_{23}$ and $^{124}Ag_{47}+^{124}Ag_{47}$ and $^{124}Pr_{59}+^{124}Pr_{59}$. We observe that\\
a. $\langle$ R $\rangle$ and $\langle$ $1/Q_{ZZ}$ $\rangle$ behave in a similar fashion. The amount of stopping decreases with the impact parameter. Reduced participant matter is the cause for this decrease.\\
b. The value of the stopping is more for $\sigma_{iso}$ than  $\sigma_{noiso}$ ($8\%$ in case of $\langle$  R $\rangle$ and $26\%$ in case of $\langle$ $1/Q_{ZZ}$ $\rangle$). This happens because, isospin dependent cross-section will lead to violent NN-collisions that will further causes the transformation of the initial longitudinal motion in other directions and hence thermalization of the system. This dominant role played by the isospin dependent cross-section gradually disappear with the impact parameter. These findings are also in supportive nature with findings of \cite{13}.\\
c. On comparing the value of stopping for both the reacting series,  $^{124}Ag_{47}+^{124}Ag_{47}$, $^{124}Pr_{59}+^{124}Pr_{59}$ and $^{40}Si_{14}+^{40}Si_{14}$, $^{40}V_{23}+^{40}V_{23}$, we found that, 
heavier masses lead to better thermalization compared to lighter nuclei.\\

\begin{figure}
\hspace{-2.0cm}\includegraphics[scale=0.45]{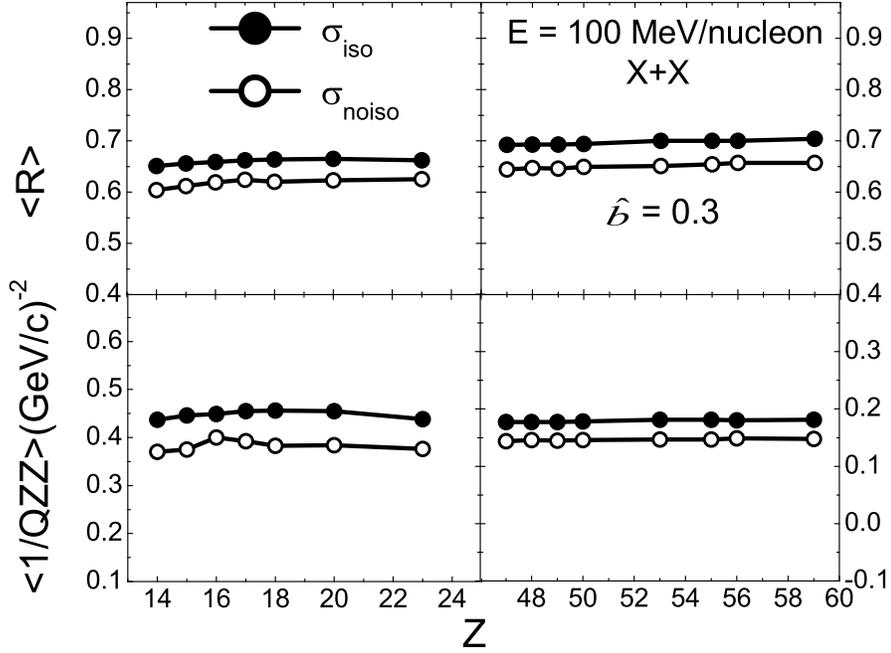}
\caption{\label{fig:3} Variation of $\langle$ R $\rangle$ and $\langle$ $1/Q_{ZZ}$ $\rangle$ with Z, in left panal for X= $^{40}V_{23}$, $^{40}Sc_{21}$, $^{40}Ca_{20}$, $^{40}Ar_{18}$, $^{40}Cl_{17}$,  $^{40}S_{16}$,  $^{40}P_{15}$, $^{40}Si_{14}$ and  in right panal for X= $^{124}Ag_{47}$, $^{124}Cd_{48}$, $^{124}In_{49}$, $^{124}Sn_{50}$, $^{124}I_{53}$, $^{124}Cs_{55}$, $^{124}Ba_{56}$, $^{124}Pr_{59}$}.
\end{figure}
As noted in the refs. \cite{13,14}, the production of free nucleons and LMF's behave in similar fashion as$\langle R \rangle$ and $\langle 1/Q_{ZZ} \rangle$.
In order to check the effect of charge asymmetry on the nuclear stopping parameters, we display in Fig.3, the variation of stopping parameters as a function of N/Z for two different cross-sections $\sigma_{iso}$ and  $\sigma_{noiso}$ for both series of reactions. In the left panels, we use $^{40}X_{m}+^{40}X_{m}$ (where m varies from 14 to 23) wheres in the right panels we use $^{124}Y_{n}+^{124}Y_{n}$ (where n varies from 47 to 59). We note that $\langle$ R $\rangle$ and $\langle$ $1/Q_{ZZ}$ $\rangle$ behave in a similar way. Further, very weak dependence is visible for charge asymmetry. This result is similar as reported in earlier figures. This observation is in agreement with the observation in ref. \cite{16}.\\ 
\begin{figure}
\hspace{-2.0cm}\includegraphics[scale=0.45]{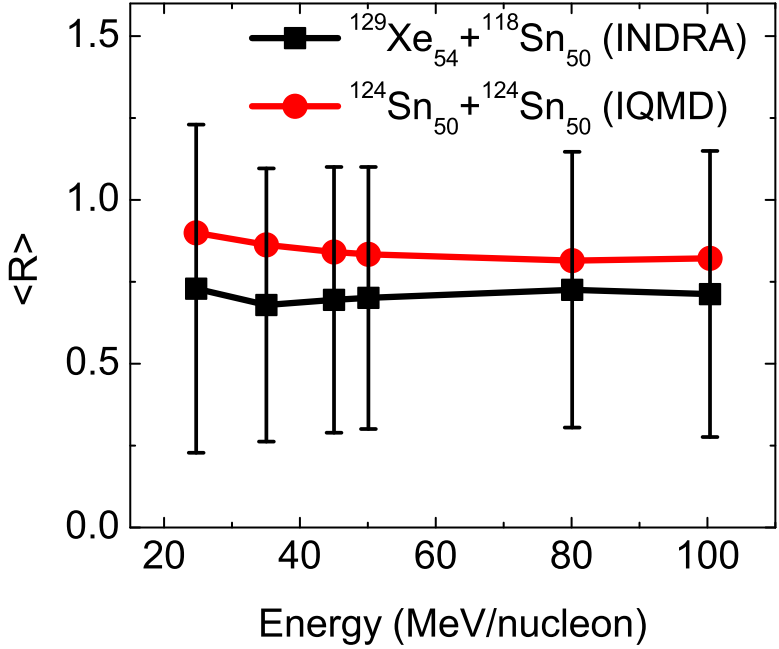}
\caption{\label{Fig:4} The anisotropy ratio $\langle$ R $\rangle$ as a function of beam energy.}
\end{figure}

To further strengthen our interpretation of the results, in fig.4, we display the comparison of theoretical results of anisotropy ratio with the experimental data obtained by the INDRA collaboration \cite{17}. Here simulations are performed for the reaction $^{124}Sn_{50}+^{124}Sn_{50}$  with $\sigma_{iso}$ reduced by $20\%$. It is worth mentioning that the results with the above choice of cross-section are in good agreement with the experimental data of ref. \cite{17}. The choice of reduced cross-section has also been motivated by ref. \cite{18} as well as many previous studies \cite{19}.
From fig.4 we also note that anisotropy ratio decreases with
increase in the incident energy. This happen because transverse
component associated with the nucleons decreases with incident energy. 
These findings are in agreement with the studies recorded in ref. \cite{13}.\\
\section{Summary}

By using the isospin dependent quantum molecular dynamics model, we have studied the effect of charge asymmetry and isospin dependent cross-section on nuclear stopping and multiplicity of free nucleons and LMF's. The calculations were carried out for $^{124}X_{m}+^{124}X_{m}$, where m varies from 47 to 59 and for $^{40}Y_{n}+^{40}Y_{n}$, where n varies from 14 to 23. Nuclear stopping as well as production of LMF's are found to depend strongly on the isospin-dependent cross-section. Moreover, theoretical results on the anisotropy ratio $\langle$ R $\rangle$ follow the same trend as recorded by INDRA collaboration.\\

{\large{\bf Acknowledgment}}

This work has been supported by a grant from the university grant commission, Government of India [Grant No. 39-858/2010(SR)].\\

\noindent


\end{document}